\documentclass[letterpaper]{article}

\usepackage{fontspec}
\defaultfontfeatures{Ligatures=TeX}
\usepackage[
    letterpaper,
    left=3.1cm,
    right=3.1cm,
    top=2.4cm,
    bottom=1.0cm,
    headheight=0pt,
    headsep=0pt,
    footskip=0.35cm
]{geometry}

\usepackage{microtype}
\usepackage{ragged2e}
\usepackage{array}
\usepackage{booktabs}
\usepackage{longtable}
\usepackage{caption}
\usepackage{changepage}
\usepackage{titlesec}
\usepackage[bottom]{footmisc}
\usepackage{fancyhdr}
\usepackage{xurl}
\usepackage{needspace}
\usepackage[hidelinks]{hyperref}
\usepackage[
    backend=biber,
    style=numeric,
    sorting=none,
    giveninits=true,
    maxbibnames=99,
    doi=true,
    url=true,
    isbn=false
]{biblatex}
\AtBeginDocument{\fontsize{11.5pt}{13.5pt}\selectfont}
\titleformat{\section}
  {\fontsize{13pt}{15pt}\selectfont\bfseries}
  {\thesection.}{0.5em}{}
\titlespacing*{\section}{0pt}{15pt}{7pt}

\titleformat{\subsection}[runin]
  {\fontsize{11.5pt}{13.5pt}\selectfont\itshape}
  {\thesubsection}{0.45em}{}
  [\textemdash\ ]
\titlespacing*{\subsection}{0pt}{9pt}{0.5em}

\titleformat{\subsubsection}[runin]
  {\fontsize{11.5pt}{13.5pt}\selectfont\itshape}
  {\thesubsubsection}{0.45em}{}
  [\textemdash\ ]
\titlespacing*{\subsubsection}{0pt}{9pt}{0.5em}

\titleformat{\paragraph}[runin]
  {\fontsize{11.5pt}{13.5pt}\selectfont\itshape}
  {}{0pt}{}
  [\textemdash\ ]
\titlespacing*{\paragraph}{0pt}{6pt}{0.5em}

\newcolumntype{P}[1]{>{\RaggedRight\arraybackslash}p{#1}}
\DeclareCaptionFont{cvaTen}{\fontsize{10pt}{13pt}\selectfont}
\makeatletter
\renewcommand\footnotesize{\@setfontsize\footnotesize{8pt}{9pt}}
\makeatother
\renewcommand{\footnoterule}{%
  \kern-3pt
  \hrule width 0.40\textwidth height 0.4pt
  \kern 2.6pt
}

\AtBeginBibliography{%
    \fontsize{10pt}{12pt}\selectfont
    \emergencystretch=2em}
\DeclareFieldFormat{labelnumberwidth}{\mkbibsuperscript{#1}}

\begin{document}
\thispagestyle{empty}

% =========================================================
% FIRST PAGE
% =========================================================

\begin{center}

{\fontsize{10pt}{13pt}\selectfont
CVA Academic Research Journal 2026
\par}

\vspace{18pt}

{\fontsize{24pt}{28pt}\selectfont
Governance Risks in Decentralized Autonomous Organizations (DAOs):\\
Implications for Traditional Financial Institutions
\par}

\vspace{14pt}

{\fontsize{13pt}{13pt}\selectfont
Christian Plöschberger\textsuperscript{1},
Irene Buller\textsuperscript{2,*},
Mark Christopher Ballandies\textsuperscript{3}
\par}

\end{center}

% Affiliations are true footnotes so they remain at the bottom of page 1.
% Affiliations are true footnotes so they remain at the bottom of page 1.

\footnotetext[1]{C.P. (christian.ploeschberger@gmail.com), CAS in Blockchain, University of Zurich, Switzerland.}

\footnotetext[2]{I.B. (irene.buller@fhnw.ch), Institute for Information Systems, University of Applied Sciences and Arts Northwestern Switzerland.}

\begingroup
\renewcommand{\thefootnote}{\fnsymbol{footnote}}
\footnotetext[1]{Corresponding author.}
\endgroup

\footnotetext[3]{M.C.B. (markchristopher.ballandies@uzh.ch), Department of Informatics (IFI), University of Zurich, Switzerland.}

\vspace{20pt}

% =========================================================
% ABSTRACT
% 10 pt / 13 pt; 1.2 cm left and right indent; <= 200 words
% =========================================================

\begin{adjustwidth}{1.2cm}{1.2cm}
\fontsize{10pt}{13pt}\selectfont
\noindent\textbf{Abstract.} As financial institutions increasingly interact with blockchain ecosystems through decentralized finance, stablecoins, tokenized assets, and distributed ledger infrastructure, they come to depend on systems governed wholly or partly by decentralized autonomous organizations (DAOs). This paper examines how DAO governance differs from traditional corporate governance and what this implies for risk management in regulated financial institutions. Using a qualitative comparative framework grounded in OECD governance principles, Basel Committee guidance, and FINMA circulars, the study maps DAO deliberation, voting, and accountability mechanisms against established governance and risk-management principles and illustrates associated risks through selected real-world cases. The analysis shows that DAOs preserve core governance functions but redistribute them across token holders, delegates, community forums, smart contracts, and legal wrapper entities. This creates distinct risks, including voting-power concentration, weak proposal controls, legal uncertainty, and unclear accountability, which established frameworks do not fully capture, while also offering two transferable strengths: a transparent, traceable decision record and a direct link between approval and execution. The paper contributes a structured assessment framework linking DAO governance characteristics to existing governance, outsourcing, technology-risk, and operational-resilience processes, enabling institutions to address DAO-governed dependencies through targeted extensions of current practices rather than a separate risk framework.
\end{adjustwidth}

\vspace{9pt}

% =========================================================
% KEY WORDS
% =========================================================

\begin{adjustwidth}{1.2cm}{1.2cm}
\fontsize{10pt}{13pt}\selectfont
\begin{center}
KEY WORDS

\vspace{6pt}

1. Decentralized Autonomous Organizations.
2. DAO Governance.
3. Financial Institutions.
4. Operational Resilience.
5. Third-Party Risk Management.
\end{center}
\end{adjustwidth}

% Resume body formatting explicitly after abstract/keywords.
\fontsize{11.5pt}{13.5pt}\selectfont

\section{Introduction}

Blockchain technology has evolved from supporting cryptocurrencies to enabling organizational structures that coordinate economic activity without centralized management \unspace\supercite{buterin2014}. Among the most prominent examples are Decentralized Autonomous Organizations (DAOs), which govern collective decisions through blockchain-based mechanisms rather than conventional corporate hierarchies \unspace\supercite{hassan2021}. Instead of relying primarily on boards, executive management and legally defined fiduciary duties, DAOs commonly use governance tokens, community proposals, off-chain discussion forums, voting processes and smart contracts that can automatically execute approved outcomes.\par
\Needspace{4\baselineskip}
This governance model matters beyond decentralized finance. Financial institutions increasingly interact with blockchain-based ecosystems through stablecoins, tokenized assets, decentralized liquidity protocols, settlement infrastructure and cryptoasset service providers. Exposure may be direct—holding or providing custody for governance tokens, offering products that reference DAO-governed protocols, using DAO-governed stablecoins, or integrating DAO-controlled protocols into treasury, collateral, settlement or liquidity-management processes—or indirect, through vendors, fintech partners, tokenization platforms, oracle providers, bridges, wallet infrastructure or custody solutions whose operations depend on DAO-governed protocols or smart-contract upgrades. Even where an institution remains legally and organizationally conventional, protocol changes, access conditions, collateral rules or upgrade decisions shaped by DAO governance can therefore affect its operational resilience, legal enforceability, client protection and third-party risk management.\par

% This governance model matters beyond decentralized finance. Financial institutions increasingly interact with blockchain-based ecosystems through stablecoins, tokenized assets, decentralized liquidity protocols, blockchain settlement infrastructure and cryptoasset service providers. Even where a regulated institution remains legally and organizationally conventional, it may depend on infrastructure whose protocol changes, access conditions, collateral rules or upgrade decisions are influenced by DAO-like governance arrangements. As a result, DAO governance can become relevant for operational resilience, legal enforceability, client protection and third-party risk management.\par

% Exposure may arise directly, for example where a bank, asset manager or market infrastructure provider holds governance tokens, provides custody for such tokens, offers products referencing DAO-governed protocols, uses DAO-governed stablecoins, or integrates DAO-controlled DeFi protocols into treasury, collateral, settlement or liquidity-management processes. It may also arise indirectly through vendors, fintech partners, tokenization platforms, oracle providers, bridges, wallet infrastructure or custody solutions whose operations depend on DAO-governed protocols or smart-contract upgrades.\par

These dependencies are relevant because decentralized finance can replicate traditional financial functions while exhibiting vulnerabilities such as operational fragilities, liquidity and maturity mismatches, leverage, interconnectedness and concentration risks \unspace\supercite{fsb2023}. The Basel Committee on Banking Supervision similarly identifies governance, technology, legal and compliance risks as important considerations for banks operating in permissionless blockchain ecosystems \unspace\supercite{bcbs2024}. More recent prudential guidance also recognizes that cryptoasset activities may expose banks to operational, market, liquidity and infrastructure risks even where institutions act as custodians or service providers rather than direct protocol participants \unspace\supercite{bcbs2026}.\par

For banks, governance is a core regulatory expectation rather than only an internal organizational matter. The Basel Committee's Corporate Governance Principles for Banks\supercite{bcbs2015} and FINMA Circular 2017/1 Corporate Governance – Banks require effective oversight, risk management, internal controls and accountability structures, while FINMA Circular 2018/3 Outsourcing and FINMA Circular 2023/1 Operational Risks and Resilience – Banks hold banks responsible for risks arising from critical third-party dependencies and technology infrastructures. When institutions rely on DAO-governed protocols or service providers, the governance arrangements of those ecosystems therefore become relevant from a prudential perspective. \par

Existing research has substantially advanced the understanding of DAO structures, voting mechanisms, legal uncertainty and socio-technical characteristics \unspace\supercite{hassan2021,ballandies2025}. Empirical studies further document concentrated tokenized voting rights, low participation, delegated voting trade-offs and centralization within prominent DAO governance systems \unspace\supercite{barbereau2023,fritsch2024,sun2024,weidener2025}. Recent taxonomy work provides a structured way to classify DAO systems across Community, Governance and Execution \unspace\supercite{anonymous2025}. The remaining analytical need is therefore not a general comparison of DAOs with corporations, but a systematic translation of DAO governance characteristics into governance and risk-management criteria relevant to regulated financial institutions.\par

This paper addresses the following research question: How do governance mechanisms in Decentralized Autonomous Organizations differ from traditional corporate governance structures, and what risk implications arise for regulated financial institutions as a result of these differences? To answer this question, the paper compares the Governance dimension of a contemporary DAO taxonomy \unspace\supercite{anonymous2025}—Deliberation, Voting and Accountability—with traditional corporate governance mechanisms derived from the OECD Principles of Corporate Governance \unspace\supercite{oecd2023}, Basel governance principles for banks \unspace\supercite{bcbs2015} and selected FINMA supervisory requirements. The objective is not to evaluate individual DAOs empirically, but to identify governance differences and risk implications at the framework level.\par

This paper makes three contributions. First, it systematically connects DAO governance taxonomy research with established corporate and banking governance principles through a functional comparison of deliberation, voting and accountability. Second, it develops an institution-oriented assessment framework that links DAO governance characteristics to existing governance oversight, third-party risk, technology-risk, compliance and operational-resilience processes. Third, it shows that DAO-governed dependencies do not require new institutional risk categories: existing risk-management structures can be extended to account for distributed decision-making, concentrated effective control and potentially limited legal accountability. The analysis addresses regulated financial institutions broadly, including banks, asset managers and market infrastructure providers, but uses banking governance as its prudential benchmark.\par

\section{Literature Review}

\subsection{Traditional Corporate Governance}

Corporate governance describes the structures and processes through which organizations are directed, controlled and held accountable. A central problem addressed by governance theory is the separation between ownership and management. Agency Theory frames this as a potential conflict between principals and agents \unspace\supercite{jensen1976}. In this paper, agency concepts provide a supplementary interpretive lens for conflicts among token holders, delegates, developers and other actors exercising effective control. The primary theoretical basis, however, is a functional governance comparison combined with an institutional risk-management benchmark rather than a separate test of agency or stakeholder theory.\par

Internationally, the OECD Principles of Corporate Governance establish widely recognized standards for transparency, shareholder rights, disclosure, board responsibilities and sustainable corporate governance \unspace\supercite{oecd2023}. For banks, Basel governance principles add a prudential perspective: the board is expected to oversee risk appetite, internal controls, compliance, culture and operational resilience \unspace\supercite{bcbs2015}. In Switzerland, FINMA Circulars 2017/1 and 2023/1 operationalize these expectations through requirements for clearly defined accountability structures, independent control functions and oversight of critical operational dependencies.\par

Traditional governance is characterized by identifiable decision-makers and enforceable responsibilities. Boards approve strategy, oversee management and supervise risk. Executive management implements decisions and remains accountable to the board. Independent control functions and internal audit provide challenge and assurance. These arrangements are usually supported by legal duties, regulatory expectations, disclosure obligations and formal reporting channels. The core governance objectives are strategic direction, accountability, transparency, oversight and risk control.\par

\subsection{DAO Governance}

DAOs represent a different organizational model \unspace\supercite{hassan2021}. They coordinate collective action using blockchain-based tools, community participation and programmable rules. Their governance often includes proposal systems, off-chain deliberation, token-based voting, delegated voting, smart contract execution and transparent on-chain records. The academic literature increasingly describes DAOs as socio-technical systems because their effectiveness depends not only on code but also on human participation, incentives, communication channels, identity mechanisms and legal structures \unspace\supercite{ballandies2025}. \par

The DAO taxonomy used in this paper organizes DAOs into three dimensions: Community (membership, interaction and interoperability), Governance (decision-making and accountability) and Execution (actions, tokens and treasury management) \unspace\supercite{anonymous2025}. This input–process–output logic explains how community input is processed through governance mechanisms and translated into outcomes. The Governance dimension, which provides the basis for comparison in this paper, consists of three categories: Deliberation refers to the creation, discussion and refinement of proposals before formal voting; Voting refers to the mechanisms through which members express preferences and approve decisions; and Accountability refers to legal entity arrangements, jurisdiction, dispute resolution and reporting mechanisms \unspace\supercite{anonymous2025}.\par

Recent empirical research shows that formal decentralization does not necessarily imply dispersed decision-making power. Barbereau et al. document highly concentrated tokenized voting rights and low exercise of those rights across DeFi projects \unspace\supercite{barbereau2023}. Fritsch et al. likewise find concentrated voting power in Compound, Uniswap and ENS, while also showing that influential addresses rarely overturn the voting preference of the wider participating community \unspace\supercite{fritsch2024}. Evidence from MakerDAO similarly identifies substantial governance centralization and a trade-off between decentralization and protocol performance \unspace\supercite{sun2024}. Delegation can mitigate participation burdens, but a recent scoping review also highlights the risk that delegated voting may reinforce concentration among influential delegates \unspace\supercite{weidener2025}. These findings indicate that concentration is an important governance vulnerability, but not by itself evidence of manipulation or governance failure. These findings concern token-governed, smart-contract-based DAOs. Bitcoin is a DAO of a different archetype: it has no legal entity and no governance token, and coordinates instead through off-chain deliberation (Bitcoin Improvement Proposals), on-chain hash-power signaling by miners and the freedom of node operators and users to adopt or reject a software version, so that governance rests on social consensus and permissionless action rather than on programmable token voting \unspace\supercite{ballandies2025bitcoin}. Such organizations widen the design space against which token-governed DAOs should be assessed. \par

\subsection{Research Gap}

The literature therefore already provides substantial evidence on DAO voting concentration, delegation, participation and governance design. It also provides taxonomies and conceptual comparisons of decentralized organizational structures. What remains less developed is an institution-oriented synthesis that connects these DAO governance mechanisms to the oversight, accountability and risk-management responsibilities of regulated financial institutions. This paper addresses that narrower gap by using deliberation, voting and accountability as a functional comparison framework and by translating the resulting governance differences into assessment criteria linked to existing governance, third-party risk, technology-risk, compliance and operational-resilience processes.\par

\section{Methodology}

\subsection{Research Design}

The study adopts a qualitative comparative design based on documentary analysis. It compares governance mechanisms rather than individual organizations, which is appropriate because the research question concerns conceptual similarities, structural differences and risk implications rather than causal measurement or statistical performance evaluation. Illustrative case studies complement the conceptual comparison for each governance category; they are not empirical tests of governance effectiveness but practical illustrations of how deliberation, voting and accountability risks have materialized in real-world DAO environments.\par

% The study adopts a qualitative comparative research design based on documentary analysis. It compares governance mechanisms rather than individual organizations. This approach is appropriate because the research question concerns conceptual similarities, structural differences and risk implications rather than causal measurement or statistical performance evaluation.\par

% To complement the conceptual comparison, the analysis also incorporates illustrative case studies for each governance category. These examples are not used as empirical tests of governance effectiveness but as practical illustrations of how risks associated with deliberation, voting and accountability have materialized in real-world DAO environments.\par

% The analysis uses secondary sources, including DAO taxonomy research, corporate governance principles and banking governance principles \unspace\supercite{anonymous2025,oecd2023,bcbs2015}.\par

\subsection{Analytical Framework}

The analytical framework is derived from the Governance dimension of the DAO taxonomy \unspace\supercite{anonymous2025}. Each DAO governance category is mapped to a traditional corporate governance equivalent and then assessed for governance risk implications. The comparison focuses on functional equivalence: whether both models perform similar governance functions, how those functions are implemented and what risks emerge from the different implementation logic.\par

\begin{table}[!htbp]
\centering
\caption{Governance Mapping Framework}\label{tab:1}
\fontsize{10pt}{11.5pt}\selectfont
\setlength{\tabcolsep}{3pt}
\renewcommand{\arraystretch}{1.08}
\begin{tabular}{@{}P{0.23\textwidth}P{0.25\textwidth}P{0.27\textwidth}P{0.18\textwidth}@{}}
\toprule
\textbf{DAO governance category} & \textbf{DAO governance attributes} & \textbf{Traditional governance equivalent} & \textbf{Primary governance objective} \\
\midrule
Deliberation & Proposal type, proposal mechanism, proposal access & Board agenda setting, executive proposals, board committees & Decision preparation \\
\addlinespace[2pt]
Voting & Voting access, voting power, voting input, aggregation & Shareholder voting, board voting, delegated authority & Decision approval \\
\addlinespace[2pt]
Accountability & Legal entity, jurisdiction, reporting, dispute resolution & Fiduciary duties, regulatory oversight, audit, courts & Responsibility and oversight \\
\addlinespace[2pt]
\bottomrule
\end{tabular}
\vspace{3pt}
{\fontsize{8pt}{11pt}\selectfont\noindent Note. DAO governance categories and attributes adapted from Makode et al.\supercite{anonymous2025}, OECD\supercite{oecd2023}, and Basel Committee on Banking Supervision\supercite{bcbs2015}.\par}
\end{table}

For each governance category, the analysis follows five steps: describe the DAO mechanism, identify the traditional equivalent, compare similarities and differences, assess governance risks from the perspective of regulated financial institutions, and map the identified risks to existing governance, third-party and outsourcing assessment where applicable, technology-risk, compliance and operational-resilience processes. This fifth synthesis step provides the methodological basis for the institutional assessment framework presented in Table~\ref{tab:5}. The framework is an analytical mapping derived from the comparison rather than an empirically validated risk-scoring instrument.\par

The DAO taxonomy of Makode et al.\supercite{anonymous2025} serves as the primary analytical framework because it classifies DAO characteristics comprehensively across Community, Governance and Execution and, within Governance, explicitly separates deliberation, voting and accountability—categories that can be compared systematically with established governance concepts. The OECD Principles of Corporate Governance \unspace\supercite{oecd2023} provide the generic benchmark for accountability, transparency, disclosure, shareholder rights and board responsibilities; the Basel Committee's Corporate Governance Principles for Banks \unspace\supercite{bcbs2015} add the prudential dimension of risk management, internal controls, compliance, operational resilience and supervisory oversight; and FINMA Circulars 2017/1, 2018/3 and 2023/1 show how these expectations are operationalized within the Swiss banking sector, including for critical third-party dependencies and technology infrastructures that may involve DAO-governed protocols \unspace\supercite{finma2016,finma2017,finma2022}. Together, these sources provide a structured basis for comparing decentralized and traditional governance while keeping the focus on the governance, accountability, operational-resilience and third-party dependency risks relevant to regulated financial institutions. \par 

\subsection{Case and Source Selection}

The cases are used as risk-illustrative examples rather than as a representative sample of DAO governance outcomes. They were selected purposively because they illuminate distinct mechanisms relevant to the analytical framework: proposal review and executable payload validation in Tornado Cash, temporary voting-power acquisition in Beanstalk, the interaction of governance tokens with market and oracle risk in Mango Markets, governance-mediated remediation in Compound, and legal and regulatory accountability in The DAO. The analysis therefore does not treat every incident as a governance attack or infer that decentralized governance is generally less effective than traditional governance.\par

The documentary analysis combines academic literature, governance and prudential standards, regulatory publications and publicly documented case materials \unspace\supercite{anonymous2025,oecd2023,bcbs2015}. Academic publications inform the conceptual comparison; OECD, Basel and FINMA documents establish institutional benchmarks; and case materials illustrate specific mechanisms or limitations. Where public case documentation does not support a causal governance claim, the analysis is limited to the governance dimension that can be substantiated.\par

\subsection{Scope and Limitations}

The paper deliberately focuses on governance mechanisms and excludes the Community and Execution dimensions of the DAO taxonomy except where they directly affect governance \unspace\supercite{anonymous2025}. It evaluates governance concepts rather than measuring governance effectiveness. The analysis addresses regulated financial institutions, but uses banking governance as its primary prudential benchmark through Basel and FINMA sources. Its implications for asset managers, market infrastructure providers and other regulated institutions should therefore be understood at the level of governance functions and risk-management principles rather than as claims of identical regulatory applicability. The illustrative cases are not a representative sample and the assessment framework is not empirically validated.\par

\section{Findings}

The following subsections analyze each category in turn and include an explicit risk assessment for regulated financial institutions.\par

\subsection{Deliberation}

In DAO governance, deliberation is the stage in which proposals are created, discussed, refined and prepared for voting. It is typically hybrid: discussion may occur off-chain in forums, Discord, GitHub or governance platforms, while approved decisions may later be encoded into on-chain proposals or executed through smart contracts. Proposal access varies. Some DAOs allow broad submission by token holders or community members, while others require thresholds, roles, reputation or delegated authority \unspace\supercite{anonymous2025}.\par

\paragraph{Traditional governance equivalent}

The corporate equivalent is structured agenda setting and board-level review. Strategic proposals usually originate from executive management and are assessed by board committees before reaching the full board. This process is narrower and more hierarchical than DAO deliberation. It emphasizes expertise, confidentiality, fiduciary responsibility and controlled information flows. By contrast, DAO deliberation emphasizes transparency, participation and collective legitimacy.\par

\begin{table}[!htbp]
\centering
\caption{Comparison of Deliberation}\label{tab:2}
\fontsize{10pt}{11.5pt}\selectfont
\setlength{\tabcolsep}{3pt}
\renewcommand{\arraystretch}{1.08}
\begin{tabular}{@{}P{0.23\textwidth}P{0.25\textwidth}P{0.27\textwidth}P{0.18\textwidth}@{}}
\toprule
\textbf{DAO governance} & \textbf{Traditional governance} & \textbf{Governance difference} & \textbf{Governance risk} \\
\midrule
Community-generated proposals & Executive-generated proposals & Decentralized agenda setting & Proposal overload \\
\addlinespace[2pt]
Public discussion forums & Confidential board discussions & Radical transparency & Information leakage \\
\addlinespace[2pt]
Token-holder participation & Committee-based expertise & Open participation & Reduced decision quality \\
\addlinespace[2pt]
Off-chain deliberation with on-chain execution & Internal deliberation followed by management implementation & Technology-supported governance & Coordination failures \\
\addlinespace[2pt]
Variable proposal access & Delegated authority & Flexible participation & Governance ambiguity \\
\addlinespace[2pt]
\bottomrule
\end{tabular}
\vspace{3pt}
{\fontsize{8pt}{11pt}\selectfont\noindent Note. Authors' comparative analysis based on the DAO governance taxonomy of Makode et al.\supercite{anonymous2025}, OECD\supercite{oecd2023}, and Basel Committee on Banking Supervision\supercite{bcbs2015}.\par}
\end{table}

\paragraph{Risk assessment}

Deliberation creates risks for financial institutions because open proposal processes can make the governance pipeline difficult to assess. High proposal volumes may produce governance fatigue and weaken decision quality. Public discussion can create information leakage where strategic, technical or economic changes are visible before they are implemented. Variable access rules can also make it unclear whether proposals reflect broad community legitimacy, concentrated influence or informal coordination by a smaller group. For institutional due diligence, the key risk indicators are proposal quality controls, participation patterns, transparency of agenda-setting, moderation or committee mechanisms, and the connection between off-chain discussion and on-chain execution.\par

\paragraph{Example}

A concrete example of deliberation risk emerged in the Tornado Cash governance incident of 2023. An attacker submitted a proposal that appeared substantially identical to a previously approved governance proposal, received community support and subsequently used hidden functionality to replace the underlying code after approval. The modified proposal ultimately granted the attacker effective control over the DAO's governance system and treasury assets \unspace\supercite{blockchainbreaches2023,copeland2023}.\par

The incident demonstrates that deliberation risk extends beyond proposal overload or limited expertise. It also includes weaknesses in technical review processes, insufficient validation that the code reviewed by participants is identical to the code ultimately executed, and inadequate governance controls capable of challenging suspicious proposals before implementation. For financial institutions, governance due diligence should therefore assess proposal-review procedures, technical audits, code-verification mechanisms, timelocks and emergency escalation processes \unspace\supercite{blockchainbreaches2023}.\par

\subsection{Voting}

Voting is the mechanism through which DAO participants approve decisions. DAO voting models vary widely \unspace\supercite{anonymous2025}. Access may be based on tokens, roles, computational power, identity or reputation. Voting power may be linear, token-weighted, delegated, quadratic or hybrid. Voting input may involve yes/no choices, ranked choices, multiple options or abstention. Aggregation may rely on simple majority, weighted majority, quorum rules or hybrid mechanisms.\par

\paragraph{Traditional governance equivalent}

Traditional corporate voting is more legally structured. Shareholders exercise voting rights through annual and extraordinary meetings, typically based on legally recognized ownership. Boards vote on strategic, risk and management decisions within a small group of identifiable individuals who owe fiduciary duties. In both systems, voting allocates decision authority. The difference lies in the source of legitimacy: corporate voting is anchored in law, ownership and fiduciary responsibility, while DAO voting is anchored in tokens, protocol participation, delegated influence or technical contribution.\par

\begin{table}[!htbp]
\centering
\caption{Comparison of Voting}\label{tab:3}
\fontsize{10pt}{11.5pt}\selectfont
\setlength{\tabcolsep}{3pt}
\renewcommand{\arraystretch}{1.08}
\begin{tabular}{@{}P{0.23\textwidth}P{0.25\textwidth}P{0.27\textwidth}P{0.18\textwidth}@{}}
\toprule
\textbf{DAO governance} & \textbf{Traditional governance} & \textbf{Governance difference} & \textbf{Governance risk} \\
\midrule
Governance token voting & Shareholder voting & Governance rights may be separated from legal ownership & Governance legitimacy risk \\
\addlinespace[2pt]
Continuous governance through ongoing proposal and voting cycles & Periodic shareholder or board votes & Higher decision frequency & Governance fatigue \\
\addlinespace[2pt]
Token-weighted voting & One-share-one-vote or board voting & Alternative allocation of influence & Whale dominance \\
\addlinespace[2pt]
Quadratic or delegated voting & Majority decision procedures & Experimental governance models & Complexity and opacity \\
\addlinespace[2pt]
Automatic execution & Management implementation after formal approval & Reduced managerial discretion & Execution errors \\
\addlinespace[2pt]
Public blockchain voting & Confidential board voting & Radical transparency & Strategic information disclosure \\
\addlinespace[2pt]
\bottomrule
\end{tabular}
\vspace{3pt}
{\fontsize{8pt}{11pt}\selectfont\noindent Note. Authors' comparative analysis based on the DAO governance taxonomy of Makode et al.\supercite{anonymous2025}, OECD\supercite{oecd2023}, and Basel Committee on Banking Supervision\supercite{bcbs2015}.\par}
\end{table}

\paragraph{Risk assessment}

Voting creates the most direct risk of governance capture. Although DAOs are often described as decentralized, voting power may be concentrated among founders, venture investors, whales or delegates \unspace\supercite{ballandies2025,barbereau2023,fritsch2024}. Low participation can magnify this effect because a small number of active voters may determine outcomes for a wider passive community. Delegated voting can reduce participation burdens, but may also concentrate practical influence among a small group of delegates \unspace\supercite{weidener2025}. Concentration should not be equated automatically with manipulation: empirical evidence shows that highly influential addresses do not necessarily overturn broader community preferences \unspace\supercite{fritsch2024}. For regulated financial institutions, voting risk assessment should therefore examine token distribution, participation history, delegation structures, quorum thresholds, actual voting concentration, proposal approval patterns and protections against manipulation. Where governance decisions are automatically executed, voting risk becomes directly connected to technology and change-management risk.\par

\paragraph{Example}

Voting-related governance risk has materialized in several major DAO incidents. Perhaps the most prominent example is the Beanstalk governance exploit of April 2022. In that case, an attacker used approximately USD 1 billion in flash-loan liquidity to acquire temporary voting power, obtain the required voting majority and approve a malicious governance proposal that ultimately diverted approximately USD 182 million from the protocol \unspace\supercite{smartcontractaudit2026}.\par

Importantly, the attack did not exploit a conventional coding vulnerability. Instead, it exploited weaknesses within the governance design itself, demonstrating how token-weighted voting systems may enable governance capture where voting power can be borrowed or accumulated temporarily without meaningful economic commitment \unspace\supercite{smartcontractaudit2026}.\par

A second example arose in Mango Markets. According to the U.S. Securities and Exchange Commission (SEC), Avraham Eisenberg manipulated the market price of MNGO, a governance token, and subsequently withdrew approximately USD 116 million from the protocol \unspace\supercite{sec2023}. The event is primarily an oracle and market-manipulation case rather than a DAO voting attack. Its relevance to this analysis is narrower: it illustrates how a governance token may simultaneously function as a voting instrument and an economic asset, creating interactions between governance influence, collateral design and market integrity.\par

\subsection{Accountability}

Accountability is the governance category where DAO governance diverges most strongly from traditional corporate governance. In DAOs, accountability is often distributed across community participants, token holders, delegates, developers, multisignature signers, legal wrapper entities and smart contracts. The DAO taxonomy identifies four accountability attributes: legal entity, jurisdiction, dispute resolution and reporting mechanisms \unspace\supercite{anonymous2025}. Early DAOs often lacked formal legal personality, while more mature DAOs increasingly use legal wrapper structures such as LLCs, foundations, associations or other entities \unspace\supercite{ballandies2025}.\par

\paragraph{Traditional governance equivalent}

In traditional corporations, responsibility is assigned to identifiable legal actors: directors, executives, auditors, shareholders and regulators. Corporate governance relies on courts, arbitration, regulators and shareholder litigation to resolve disputes. Reporting is formalized through periodic reports, regulated disclosures and audit processes. These arrangements create a clearer chain of responsibility than is typically present in decentralized governance systems.\par

\paragraph{Comparative analysis}

DAO accountability can improve transparency because proposals, votes, treasury transactions and smart contract interactions may be visible. However, transparency does not automatically create accountability. Observability shows what happened, but not necessarily who is legally or operationally responsible. Jurisdiction also remains complex because participants, smart contracts, treasury assets and users may be globally distributed.\par

\begin{table}[!htbp]
\centering
\caption{Comparison of Accountability}\label{tab:4}
\fontsize{10pt}{11.5pt}\selectfont
\setlength{\tabcolsep}{3pt}
\renewcommand{\arraystretch}{1.08}
\begin{tabular}{@{}P{0.23\textwidth}P{0.25\textwidth}P{0.27\textwidth}P{0.18\textwidth}@{}}
\toprule
\textbf{DAO governance} & \textbf{Traditional governance} & \textbf{Governance difference} & \textbf{Governance risk} \\
\midrule
Legal wrapper optional or evolving & Legal entity mandatory & Legal personality may be unclear & Legal uncertainty \\
\addlinespace[2pt]
Community accountability & Directors and executives accountable & Distributed responsibility & Weak individual accountability \\
\addlinespace[2pt]
Blockchain transparency & Periodic reporting & Continuous disclosure & Information overload \\
\addlinespace[2pt]
Smart contract execution & Management execution & Technology replaces discretion & Automated failures \\
\addlinespace[2pt]
Forking or decentralized arbitration & Courts and regulators & Alternative dispute resolution & Enforcement uncertainty \\
\addlinespace[2pt]
Global participation & National legal jurisdiction & Cross-border governance & Regulatory fragmentation \\
\addlinespace[2pt]
\bottomrule
\end{tabular}
\vspace{3pt}
{\fontsize{8pt}{11pt}\selectfont\noindent Note. Authors' comparative analysis based on the DAO governance taxonomy of Makode et al.\supercite{anonymous2025}, OECD\supercite{oecd2023}, and Basel Committee on Banking Supervision\supercite{bcbs2015}.\par}
\end{table}

\paragraph{Risk assessment}

Regulated financial institutions need identifiable counterparties, enforceable rights, escalation paths and accountable decision-makers. DAO governance may leave uncertainty over legal personality, jurisdiction, liability, fiduciary duties and dispute resolution; legal wrappers partially mitigate these risks but may not align formal legal responsibility with actual governance influence. Institutional assessment should therefore examine the existence and role of legal entities, governing law, enforceability of claims, responsible administrators or multisignature signers, upgrade rights, emergency powers, reporting practices and dispute mechanisms. Where these elements are weak or unclear, reliance on the DAO-governed ecosystem creates elevated legal, operational and regulatory risk. This contrasts with FINMA Circular 2017/1, which requires clear allocation of responsibilities, effective oversight mechanisms and documented accountability structures for material activities and risks; the absence of clearly accountable legal or natural persons complicates the governance assessments regulated institutions must perform.\par

\paragraph{Example}

Accountability failures can become particularly significant when governance-approved changes create material consequences without a clearly identifiable actor capable of rapid intervention. Compound's Proposal 62 provides an illustrative example. Following approval of a governance proposal in 2021, a bug in the upgraded Comptroller contract resulted in the unintended distribution of large quantities of COMP governance tokens. The potential impact was estimated at approximately USD 80–90 million \unspace\supercite{theblock2021,claburn2021}.\par

Because governance authority had been intentionally decentralized and immediate administrative intervention capabilities were limited, remediation required progression through the protocol's governance process rather than direct managerial action. This illustrates a recurring DAO governance challenge: mechanisms designed to reduce centralized control may complicate incident response because remediation often depends on governance processes rather than unilateral managerial intervention. The resulting response speed may vary considerably across DAOs.\par

A related accountability example can be found in the original 2016 DAO incident. The SEC reported that an attacker exploited a flaw in The DAO's code and diverted roughly one-third of the organization's assets before any investment projects had commenced \unspace\supercite{sec2017a}. Beyond the financial losses, the subsequent SEC investigation highlighted broader questions concerning investor protection, legal responsibility and the regulatory status of DAO governance structures \unspace\supercite{sec2017a,sec2017b}.\par

\subsection{Summary of Findings}

The comparison yields three category-specific findings. Deliberation shifts governance risk toward proposal initiation, agenda-setting influence, proposal quality and the adequacy of technical and procedural review; open participation and public discussion do not by themselves ensure that consequential proposals receive effective scrutiny before execution. Voting shifts attention toward the distribution of effective control, where token concentration, low participation and delegation can allow a small group of active participants to determine outcomes, although concentration must be assessed together with actual voting behavior and safeguards \unspace\supercite{barbereau2023,fritsch2024,weidener2025}. Accountability represents the largest structural divergence: whereas traditional institutions operate within established structures of legal responsibility, supervisory oversight and enforceable organizational duties, DAO governance may distribute consequential authority among actors whose responsibilities and legal relationships are far less clearly defined, to an extent that depends on the DAO’s actual legal and organizational structure.\par

Taken together, these findings show that DAO governance does not eliminate governance functions; it redistributes decision-making authority, control and accountability across different actors, technical mechanisms and legal arrangements \unspace\supercite{hassan2021,anonymous2025}. Existing governance principles therefore remain relevant, but their application must be adapted to decentralized, hybrid and technology-mediated dependencies.\par

\section{Discussion}

\subsection{Governance Functions and Institutional Risk}

The findings extend the existing DAO governance literature by connecting observed governance mechanisms to the oversight responsibilities of regulated financial institutions. Prior studies document concentrated voting rights, low participation, delegation trade-offs and protocol-specific centralization \unspace\supercite{barbereau2023,fritsch2024,sun2024,weidener2025}. The institutional significance of these characteristics, however, depends on how they affect effective control, the reliability of decision-making and the enforceability of responsibility.\par

This distinction matters because formal voting concentration is not sufficient evidence of governance failure. Institutional assessment must consider how governance rights are exercised, which safeguards constrain consequential decisions, whether technical execution can be delayed or challenged, and whether escalation or remediation mechanisms are available. Agency concepts remain useful here because formal ownership, delegated authority and actual influence may diverge, while the incentives of decision-makers may not align with those of protocol users or institutions exposed to the protocol \unspace\supercite{jensen1976}.\par

From a risk-management perspective, the findings support adapting established institutional frameworks rather than creating a separate DAO-specific risk taxonomy. DAO governance does not create entirely new categories of risk; it changes how familiar governance, operational, legal and technology risks manifest. Governance oversight, legal and compliance assessment, technology risk management and operational-resilience processes therefore remain relevant, but require additional indicators concerning proposal controls, effective voting influence, delegated authority, smart-contract execution and identifiable accountability arrangements.\par

%From a risk-management perspective, the findings support adapting established institutional frameworks rather than replacing them with a separate DAO-specific risk taxonomy. Governance oversight, legal and compliance assessment, technology risk management and operational-resilience processes remain relevant, but their application may require additional indicators concerning proposal controls, effective voting influence, delegated authority, upgrade mechanisms and identifiable accountability arrangements.\par

\subsection{Accountability as the Most Pronounced Governance Challenge}

Accountability is the most difficult area for financial institutions. DAO transparency improves observability, yet transparency is not enforceability: a public record does not by itself identify a legally responsible actor, create fiduciary duties or ensure access to courts or regulators. This creates an accountability paradox: DAOs may be more transparent than corporations while remaining less clear about responsibility. Legal wrapper structures (foundations, LLCs, associations or similar entities) partially address this challenge by creating points of legal recognition, contractual capacity and regulatory engagement \unspace\supercite{ballandies2025}. However, they often coexist with decentralized governance rather than replacing it, so institutions must assess both the formal legal entity and the practical governance system that can influence protocol behavior.\par

The absence of a legal entity is not necessarily a deficiency. In Bitcoin, the lack of any legal person able to admit, expel or direct participants is what guarantees permissionless participation and individual autonomy, whereas a legal wrapper binds a DAO to a jurisdiction at the cost of that autonomy \unspace\supercite{ballandies2025bitcoin}. For institutions this is a trade-off: a wrapper offers a counterparty and an escalation path, but if it does not control protocol behavior it provides accountability only in form. Assessments should therefore establish what the legal entity can actually decide and treat DAOs that deliberately operate without one as a distinct archetype rather than as immature versions of wrapped DAOs.\par

% Accountability is the most difficult area for financial institutions. DAO transparency can improve observability: proposals, votes, treasury movements and smart contract actions may be visible. Yet transparency is not equivalent to enforceability. A public record does not by itself identify a legally responsible actor, create fiduciary duties or ensure access to courts or regulators. This creates an accountability paradox: DAOs may be more transparent than corporations while remaining less clear about responsibility.\par

% Legal wrapper structures partially address this challenge \unspace\supercite{ballandies2025}. Foundations, LLCs, associations or similar entities can create points of legal recognition, contractual capacity and regulatory engagement. However, they often coexist with decentralized governance rather than replacing it. Financial institutions must therefore assess both the formal legal entity and the practical governance system that can influence protocol behavior.\par

\subsection{Implications for Financial Institutions}
The preceding analysis identifies four areas in which financial
institutions can translate DAO governance risks into institutional
practice.

\subsubsection{Governance Due Diligence}
For regulated financial institutions, DAO governance risk should
be incorporated into existing operational resilience, outsourcing,
technology risk management and third-party risk management frameworks. The relevant question is not only whether an institution directly holds cryptoassets or governance tokens, but also whether critical business processes depend on infrastructure whose governance is controlled, wholly or partially, through decentralized decision-making mechanisms \unspace\supercite{bcbs2015,fsb2023}.\par

%From a risk-management perspective, DAO governance does not necessarily create entirely new categories of risk. Rather, it changes how familiar governance, operational, legal and technology risks manifest. Consequently, existing governance and third-party risk management frameworks remain relevant, but they must be expanded to account for decentralized decision-making structures, token-based voting systems, smart-contract execution and distributed accountability mechanisms.\par

Incorporating DAO governance into existing frameworks is consistent with prudential expectations concerning external dependencies and operational resilience. FINMA Circular 2018/3 Outsourcing addresses the management of qualifying outsourcing relationships, while FINMA Circular 2023/1 emphasizes the resilience of services supporting critical functions. Dependence on DAO-governed infrastructure should not automatically be classified as regulated outsourcing; the legal characterization depends on the institution, the function, the contractual relationship and the applicable supervisory perimeter. Regardless of classification, governance arrangements may become a relevant object of due diligence where they can materially affect a service on which the institution depends. Table~\ref{tab:5} maps DAO governance characteristics to the existing governance, outsourcing, compliance and operational-resilience processes through which such assessments can be conducted.\par

\begin{center}
\fontsize{10pt}{11.5pt}\selectfont
\setlength{\tabcolsep}{3pt}
\renewcommand{\arraystretch}{1.08}
\begin{longtable}{@{}P{0.205\textwidth}P{0.205\textwidth}P{0.29\textwidth}P{0.205\textwidth}@{}}
\caption{DAO Governance Assessment Framework for Financial Institutions}\label{tab:5}\\
\toprule
\textbf{DAO governance characteristic} & \textbf{Potential institutional impact} & \textbf{Governance assessment questions} & \textbf{Existing risk management process} \\
\midrule
\endfirsthead
\multicolumn{4}{l}{\fontsize{10pt}{11.5pt}\selectfont \textit{Table 5 continued}}\\
\toprule
\textbf{DAO governance characteristic} & \textbf{Potential institutional impact} & \textbf{Governance assessment questions} & \textbf{Existing risk management process} \\
\midrule
\endhead
\midrule
\multicolumn{4}{r}{\fontsize{8pt}{9pt}\selectfont Continued on next page}\\
\endfoot
\bottomrule
\endlastfoot
Decentralized proposal processes & Difficulty assessing governance maturity and change initiation & Who can submit proposals? Are proposals independently reviewed and technically validated? & Third-party governance review \\
\addlinespace[2pt]
Token-based voting & Concentrated governance influence and governance capture risk & How concentrated is voting power? Are anti-manipulation controls effective? & Ownership and control assessment \\
\addlinespace[2pt]
Continuous governance & Frequent protocol changes & How often are governance proposals approved? Can changes be monitored before implementation? & Change management \\
\addlinespace[2pt]
Smart-contract execution & Automatic implementation of governance decisions & Are timelocks, veto rights or emergency pause mechanisms available? & Technology risk management \\
\addlinespace[2pt]
Distributed accountability & Unclear responsible counterparties & Is there a legal entity? Who can represent the protocol legally? & Legal and outsourcing due diligence \\
\addlinespace[2pt]
Global governance participation & Regulatory and jurisdictional uncertainty & Which law applies? Which entities can be subject to regulatory oversight? & Compliance and jurisdiction assessment \\
\addlinespace[2pt]
Privileged governance powers & Key-person and operational resilience risks & Who controls multisignature wallets, admin keys or emergency governance functions? & Operational resilience and cyber risk management \\
\addlinespace[2pt]
Governance failure or protocol compromise & Service disruption and vendor dependency risk & Can the institution exit the dependency if governance quality deteriorates? & Exit planning and business continuity \\
\addlinespace[2pt]
\end{longtable}
\end{center}
\vspace{-6pt}
{\fontsize{8pt}{11pt}\selectfont\noindent Note. Author's analysis based on OECD\supercite{oecd2023}, Basel Committee on Banking Supervision\supercite{bcbs2015,bcbs2024,bcbs2026}, and the governance risks identified throughout this paper.\par}
\vspace{6pt}

The framework shows that governance assessment must extend beyond traditional vendor due diligence. Institutions should determine whether vendors depend on DAO-governed smart contracts, stablecoins, bridges, oracle networks, tokenized asset platforms or liquidity protocols, and whether governance decisions can alter collateral requirements, transaction fees, access controls, liquidation parameters, redemption mechanisms or service availability without contractual negotiation or regulatory approval \unspace\supercite{bcbs2024,bcbs2026}. Governance quality deserves particular attention: who may submit proposals, how proposals are reviewed, whether payloads are technically validated prior to execution, whether emergency veto mechanisms exist, and whether governance capture techniques such as flash-loan voting or vote buying are effectively mitigated. The Beanstalk and Tornado Cash incidents show how weaknesses in proposal review, voting safeguards and execution controls can turn governance processes themselves into attack vectors \unspace\supercite{blockchainbreaches2023,copeland2023,smartcontractaudit2026}.\par

% The framework demonstrates that governance assessments must extend beyond traditional vendor due diligence. Institutions should determine whether vendors depend on DAO-governed smart contracts, stablecoins, bridges, oracle networks, tokenized asset platforms or liquidity protocols. They should additionally assess whether governance decisions can alter collateral requirements, transaction fees, access controls, liquidation parameters, redemption mechanisms or service availability without contractual negotiation or regulatory approval \unspace\supercite{bcbs2024,bcbs2026}.\par

% Particular attention should be paid to governance quality. Relevant questions include who may submit proposals, how governance proposals are reviewed, whether proposal payloads are technically validated prior to execution, whether emergency veto mechanisms exist, whether voting power is concentrated among a small number of token holders and whether governance capture techniques such as flash-loan voting or vote buying are effectively mitigated. The Beanstalk and Tornado Cash incidents demonstrate how weaknesses in proposal review, voting safeguards and execution controls can transform governance processes themselves into attack vectors \unspace\supercite{blockchainbreaches2023,copeland2023,smartcontractaudit2026}.\par

Institutions should also assess accountability structures. Due diligence should address the existence of legal wrapper entities, applicable jurisdiction and governing law, dispute-resolution mechanisms, contractual enforceability, emergency administrative powers and the identities of individuals or entities controlling multisignature wallets or privileged governance functions. The DAO investigation and the Mango Markets enforcement action illustrate that governance failures may evolve into broader legal, regulatory, securities-law and market-conduct concerns \unspace\supercite{sec2017a,sec2017b,sec2023}.\par

\subsubsection{Incident Management and Operational Resilience}

Beyond ex-ante assessment, institutions must evaluate how DAO-governed dependencies behave during operational disruptions, cyber incidents and governance failures. Whereas traditional outsourcing arrangements provide identifiable management structures and contractual escalation pathways, DAO-governed ecosystems rely on a combination of governance delegates, multisignature signers, foundations, protocol developers, security councils and community participants whose responsibilities are not always formally documented. Incident playbooks should therefore identify the parties capable of influencing remediation outcomes, including delegates, protocol maintainers, foundation representatives, multisignature signers, bridge operators, oracle providers and relevant cryptoasset service providers, and should determine whether a governance proposal, smart-contract upgrade, voting event, oracle change, delegation change or privileged governance action contributed to an incident. Institutions should further evaluate whether emergency intervention mechanisms exist, including pause functions, veto rights, challenge periods, timelocks, emergency committees or multisignature approval controls.\par

In DAOs governed by social consensus rather than token voting, the institution itself may be one of these parties: large custodians and ETF issuers belong to the economic majority whose choice of software version or post-fork chain can decide protocol outcomes, and the prospectus of the largest Bitcoin ETF already reserves the issuer's discretion over which chain the fund follows after a hard fork \unspace\supercite{ballandies2025bitcoin}. Due diligence should therefore also cover the institution's own influence, since concentrated institutional holdings reduce network decentralization and shift effective, if not formal, accountability for protocol outcomes onto the institution.\par

The governance delays observed in Compound's Proposal 62 demonstrate that decentralization can complicate remediation when rapid intervention is required: while decentralized governance reduces certain forms of unilateral administrator risk, it can also slow corrective measures when governance-approved changes produce unintended consequences \unspace\supercite{claburn2021,theblock2021}. The incident examples in this paper should be read as illustrations of such decision-making, accountability and remediation challenges rather than as assessments of the responsiveness of individual DAOs. Institutions should consequently evaluate governance not only under normal operating conditions but under stress: who can authorize emergency actions, how quickly vulnerabilities can be addressed and whether accountability remains sufficiently clear during incidents affecting clients, operations or regulatory obligations. This aligns with FINMA Circular 2023/1, whose emphasis on preparedness for disruption, effective escalation and the ability to restore critical services under stressed conditions extends, for DAO-governed dependencies, to governance processes, voting mechanisms and emergency intervention rights.\par

\subsubsection{Governance Risk Appetite and Compensating Controls}

Where DAO-governed infrastructure supports critical or important business functions, governance considerations should become part of the institution's risk appetite framework. As with traditional outsourcing, institutions should define acceptable thresholds for governance-related risk indicators and escalation criteria where thresholds are exceeded. Relevant indicators include voting-power concentration, proposal participation rates, quorum levels, timelock duration, legal enforceability, administrator transparency, multisignature governance design, smart-contract audit coverage and incident-response maturity. Monitoring should also track whether governance practices and control mechanisms remain adequate as protocols grow in scale and complexity.\par

Where governance arrangements do not satisfy institutional expectations, compensating controls may include exposure limits, enhanced governance monitoring, independent smart-contract reviews, additional contractual protections through legal wrapper entities, alternative service providers, contingency arrangements and exit strategies designed to reduce dependency on a single DAO-governed protocol.\par

Ultimately, DAO governance should be viewed as a component of governance risk, operational resilience, outsourcing oversight and third-party risk management rather than as a technology-specific concern, because governance decisions in DAO-governed ecosystems can directly affect the availability, integrity and legal certainty of services on which financial institutions depend \unspace\supercite{bcbs2024,bcbs2026,finma2017,finma2022}. As financial institutions deepen their interactions with blockchain ecosystems, the ability to assess governance quality, accountability structures and governance resilience may become as important as the assessment of financial strength, cybersecurity capabilities and legal enforceability in traditional vendor relationships.\par

\subsubsection{Secondary Insight: Selective Adoption of DAO Governance Mechanisms}

Although the primary focus of this paper is governance risk, selected DAO mechanisms provide two secondary insights for traditional institutions. The first is procedural transparency: because proposals, deliberation, votes, treasury movements and executed changes are publicly recorded, DAO governance produces a continuous, time-stamped record of how decisions are prepared, approved and implemented—a level of traceability that periodic corporate reporting rarely achieves. This record does not substitute for accountability, as Section~5.2 makes clear, but as a governance audit trail it is a strength in its own right. The second is the technical linkage between authorized decisions and implementation, which can strengthen implementation discipline. Any institutional adoption of either would need to remain compatible with regulatory oversight, fiduciary duties and clear management accountability \unspace\supercite{hassan2021,ballandies2025,oecd2023,bcbs2015}.\par

Of the two, the more directly transferable is the second: the connection between governance decisions and their execution. Once predefined governance conditions have been satisfied, approved changes can be implemented automatically through smart contracts after an appropriate review period. This minimizes manual intervention and reduces the risk that implementation diverges from the approved decision \unspace\supercite{buterin2014,anonymous2025,ballandies2025}.\par

Although regulated financial institutions cannot fully automate governance decisions without human oversight, many post-approval activities could be translated into controlled workflow automation. Examples include the automatic publication of approved policies, initiation of mandatory employee attestations, activation of technical controls, assignment of implementation tasks, and continuous monitoring of completion. Such approaches have the potential to reduce operational risk, improve consistency, and shorten implementation cycles while preserving management accountability \unspace\supercite{bcbs2015,finma2016,finma2022}.\par

\section{Conclusion}

This paper examined how governance mechanisms in DAOs differ from traditional corporate governance structures and what this implies for regulated financial institutions. Using the Governance dimension of a contemporary DAO taxonomy \unspace\supercite{anonymous2025} as the analytical framework, it compared DAO deliberation, voting and accountability mechanisms with governance principles derived from the OECD Principles of Corporate Governance \unspace\supercite{oecd2023}, the Basel Committee's Corporate Governance Principles for Banks \unspace\supercite{bcbs2015} and FINMA supervisory requirements.\par

%This paper examined how governance mechanisms in Decentralized Autonomous Organizations (DAOs) differ from traditional corporate governance structures and assessed the resulting implications for regulated financial institutions. Using the Governance dimension of a contemporary DAO taxonomy \unspace\supercite{anonymous2025} as the analytical framework, the study compared DAO deliberation, voting and accountability mechanisms with governance principles derived from the OECD Principles of Corporate Governance \unspace\supercite{oecd2023} and the Basel Committee's Corporate Governance Principles for Banks \unspace\supercite{bcbs2015}.\par

The analysis shows that DAO governance does not eliminate traditional governance functions but redistributes them across token holders, delegates, community forums, off-chain platforms, smart contracts and, increasingly, legal wrapper entities. Deliberation still prepares decisions, voting still allocates decision authority and accountability still provides legitimacy and oversight, but through architectures that differ significantly from the hierarchical, legally embedded structures of regulated institutions. Each category carries distinct risks: deliberation mechanisms may weaken proposal quality and review; voting mechanisms may enable governance capture, concentration of influence and manipulation of governance rights; and accountability mechanisms—the largest divergence identified—may provide full transparency without clear legal responsibility, enforceable rights or identifiable counterparties.\par

% The analysis demonstrates that DAO governance does not eliminate traditional governance functions. Rather, it redistributes them across a combination of token holders, delegates, community forums, off-chain platforms, smart contracts and, increasingly, legal wrapper entities. Deliberation continues to perform the function of decision preparation, voting remains the primary mechanism for allocating decision authority and accountability continues to provide legitimacy and oversight. However, these functions are implemented through governance architectures that differ significantly from the hierarchical and legally embedded structures typically found in regulated financial institutions.\par

% The comparative assessment identified distinct governance risks associated with each governance category. Deliberation mechanisms may introduce risks related to proposal quality, information asymmetries and weaknesses in proposal review processes. Voting mechanisms may create vulnerabilities to governance capture, concentration of influence, low participation and manipulation of governance rights. Accountability emerged as the most pronounced governance challenge identified in the comparison because transparency of decision-making does not necessarily translate into clear legal responsibility, enforceable rights or identifiable counterparties. The analysis therefore suggests that accountability represents the largest divergence between DAO governance and traditional corporate governance.\par

The illustrative cases show how these vulnerabilities and remediation constraints materialize in practice, but they are deliberately risk-illustrative rather than representative. Beanstalk and Tornado Cash involve direct governance-control weaknesses, whereas Mango Markets is primarily an oracle and market-manipulation case, Compound illustrates governance-mediated remediation after a software defect, and The DAO illustrates a code exploit with subsequent legal and regulatory consequences. Their diversity underlines that governance risks vary materially across DAOs, and that increasingly mature DAOs have adopted legal wrappers, delegated governance, security councils, timelocks and other mechanisms to strengthen accountability. Governance assessments should therefore focus on the characteristics of individual DAO governance systems rather than treating decentralization itself as a uniform source of risk, and should recognize that DAOs governed by social consensus without a legal entity or governance token, of which Bitcoin is the archetype, form a distinct category rather than a deficient one \unspace\supercite{ballandies2025bitcoin}. \par

For regulated financial institutions, DAO governance should be treated as a component of governance risk, operational resilience, outsourcing oversight and third-party risk management. The primary challenge is not the creation of new governance frameworks but the adaptation of existing ones to environments in which authority is distributed across decentralized actors and programmable systems. The primary contribution of this paper is the structured comparison and assessment framework that connects DAO taxonomy research with corporate governance, banking governance and operational-resilience principles; it provides a practical basis for evaluating governance concentration, proposal controls, accountability structures, legal enforceability, smart-contract execution mechanisms, incident-response capabilities and governance-related operational dependencies within existing risk-management practices.\par

DAO governance should not be viewed only as a source of risk. Two transferable strengths emerge from the comparison. The first is procedural transparency: because proposals, deliberation, votes, treasury movements and executed changes are publicly recorded, DAO governance produces a continuous and verifiable audit trail of how decisions are prepared, approved and implemented—a level of traceability that periodic corporate reporting rarely achieves. The second is the direct link between governance approval and execution: while banks must preserve human oversight, fiduciary responsibility and formal accountability, DAO-inspired post-approval workflow automation, combined with time-stamped decision records, could strengthen implementation discipline and narrow the gap between formal governance decisions and operational execution.\par

Future research could examine governance effectiveness across DAO maturity levels, empirically assess governance outcomes in specific protocols and investigate how emerging regulatory frameworks influence the convergence of decentralized and traditional governance models, contributing to a more nuanced understanding of how governance can be designed, supervised and assessed in increasingly interconnected financial and blockchain-based ecosystems.\par

% =========================================================
% REFERENCES
% CVA template uses superscript numbered references/endnotes.
% =========================================================
\section*{References}
\printbibliography[heading=none]

\section*{Author Contributions}

CP: Conceptualization, investigation, methodology, and writing---original draft.
IB: Literature review and positioning, manuscript coherence, contribution framing, consistency review across the abstract, findings, discussion, and conclusion, and writing---review and editing.
MCB: Supervision, domain expertise, methodological guidance, conceptual development, and writing---review and editing.
All authors contributed to the interpretation and refinement of the study and approved the final manuscript.

\end{document}